\documentstyle[prl,aps,epsfig,floats]{revtex}

\newcommand\beq{\begin{equation}}
\newcommand\eeq{\end{equation}}
\newcommand\bea{\begin{eqnarray}}
\newcommand\eea{\end{eqnarray}}

\newcommand\bi{\begin{itemize}}
\newcommand\ei{\end{itemize}}

\begin{document}

\draft

\textheight=23.8cm
\twocolumn[\hsize\textwidth\columnwidth\hsize\csname@twocolumnfalse\endcsname

\title{\Large Effects of interaction on an adiabatic quantum electron pump}
\author{\bf Sourin Das and Sumathi Rao } 

\address{\it 
Harish-Chandra Research Institute,
Chhatnag Road, Jhusi, Allahabad 211019, India}
\date{\today}
\maketitle

\begin{abstract}

We study the effects of inter-electron interactions on the charge
pumped through an adiabatic quantum electron pump. The pumping
is through a system of barriers, whose
heights are deformed adiabatically. (Weak) interaction effects are introduced
through a renormalisation group flow of the scattering matrices
and the pumped charge is shown to {\it always} approach a quantised value
at low temperatures or long length scales.  
The maximum value of the pumped charge is set by
the number of barriers and is given by $Q_{\rm max} = n_b -1$.
The correlation between the transmission 
and the charge pumped is studied by seeing how much of the
transmission is enclosed by the pumping contour.  The (integer)
value of the pumped charge at low temperatures is determined by
the number of transmission maxima enclosed by the pumping contour.
The dissipation at finite  temperatures leading to 
the non-quantised values of the pumped charge scales as a power law with the
temperature ($Q-Q_{\rm int} \propto T^{2\alpha}$),
or with the system size  ($Q-Q_{\rm int} \propto L_s^{-2\alpha}$),
where $\alpha$ is a measure
of the interactions and vanishes at $T=0 ~(L_s=\infty)$.
For a double barrier system, our 
result agrees with the quantisation of pumped charge seen in 
Luttinger liquids.

\end{abstract}


]

In the last few years, there has been a lot of interest in the phenomenon 
of electron transport at zero bias
via the mechanism of adiabatic pumping. The
parameters of a system are slowly varied as a function of time, such that
they return to themselves after each cycle, and the net
result is charge transport\cite{ALTGLAZ}. 
The idea of quantised charge pumping was first introduced by 
Thouless\cite{THOULESS} for a spatially periodic system, where the
quantisation could be proved using topological arguments. This
was further investigated theoretically in Refs.\cite{NIU}.

Experimentally, the possibility of transferring an integer number
of charges through an unbiased system has been achieved in small
semi-conductor dots in the Coulomb blockade (CB) regime\cite{KOUW,POTHIER}.
It has also been achieved in surface-acoustic-wave based 
devices\cite{SHILTON}, but here too, the quantisation has been
attributed to CB. An adiabatic electron pump
that produces a dc voltage, in response to a cyclic change in the confining
potential of an {\it open} quantum dot has also been experimentally
constructed\cite{SWITKES}; however, here the charge is not quantised. 

The {\it open} dot quantum pumps, where the quantum interference of the
wave-function, rather than the Coulomb blockade, plays a dominant
role, has been of recent theoretical interest.  A scattering
approach to such a pump was pioneered by Brouwer\cite{BROUWER}
and others
\cite{SPIVAK,SHUTENKO}, who, building on earlier results\cite{BPT},
related the pumped current to derivatives of the the scattering
matrix. Using this approach, there have been several recent theoretical
studies
\cite{BURM,AVR,LEV,ENT,WEI,ZH,PB,REN,AHAR,BLAA,CB,PVB,B,U,ZHU,AO,SB,C,AB,SELA}.
The conditions under which the charge is (almost) quantised in
open quantum dots, has been explored in detail in 
Refs.\cite{ALEINER,EWAA1,EWAA2,SHARMA}. 
Ref.\cite{ALEINER} shows that the pumped charge is
produced by both nondissipative and dissipative currents. It also  
shows that charge quantisation is achieved when the dissipative
conductance vanishes. This is also emphasized in Ref.\cite{SHARMA}, where
the adiabatic charge transport in a Luttinger liquid has been
studied.  In Refs\cite{EWAA1,EWAA2},
the  correlation between the conditions for resonant
transmission  and quantised charge has been emphasized. 

In this
paper, we include the effects of inter-electron interactions along
with the interference effects, using a renormalisation group (RG)
method developed recently\cite{YUE,LAL,POLYAKOV,DRS}
for the flow of the scattering matrices. 
We show that at low temperatures, the inclusion
of interactions, leads to  quantisation of the pumped charge.
As seen earlier in Refs.\cite{ALEINER} and \cite{SHARMA},
we find that the charge pumped is a sum of two terms - a quantised
term, and a dissipative term. The quantised term, which is topological
in nature and is due to quantum interference is always present.
For non-interacting electrons, the dissipative term is also large
and masks the quantisation due to the first term. However, the RG  flow of the
$S$-matrices, due to the interaction, causes the dissipative
term to vanish at long length scales or low
temperatures making
the quantisation of the pumped charge visible.

\begin{figure}[htb]
\begin{center}
\epsfig{figure=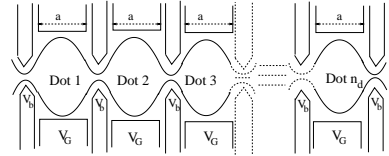,width=5.0cm}
\end{center}
\caption{Schematic diagram of a multiple dot
system ($n_d$ dots or $n_b+1$ barriers) 
defined on a two dimensional electron gas.  
The barriers forming the dots are denoted as $V_b$ and the gate 
voltages controlling the density in the dots are denoted as $V_G$.}
\end{figure}

The quantum pump is formed by a system of coupled quantum dots 
(shown in Fig.1), where the barriers forming the dot are 
periodically  modulated as 
\bea
V_i &\equiv& V_1 = V_0 + V_p \cos (\omega t), 
~i \le n_b/2  ~ {\rm for}~ n_b= {\rm even}, \nonumber \\ 
&&\quad i \le n_b/2+1 ~{\rm for}~ n_b={\rm odd}, \nonumber \\
V_i &\equiv& V_2= V_0 + V_p \cos (\omega t + \delta), ~{\rm for~
the ~remaining} .
\label{one}
\eea
Clearly, for $n_b$ barriers, we have $n_d=n_b-1$ dots.
Here $\omega$ is related to the time period as 
$\omega =\tau/2\pi$ and $\delta$ is the phase difference 
between the two time-varying potentials.
Since this potential breaks the parity symmetry, it allows the shape of 
all the dots to be varied. As has been emphasized earlier, this
is a necessary condition to get non-zero pumped charge. 

Although we use $\delta$-function barriers,
we expect the calculations  to be robust to changing the
form of the barriers.
Following the work of Ref.\cite{NT}, we also expect these results
to be robust to weak disorder. 
We treat the
dot within a one-dimensional effective Hamiltonian\cite{MATVEEV}, since
they are coupled to the leads by a single
channel quantum point contact.
(Although in real quantum dots, transmission properties can  
fluctuate and energy level spacings can vary and can
even be chaotic, this modelling works to understand simple
physical features of higher dimensional  quantum dots.) 
The width of the dots (effectively the width of the quantum well that
we use) is given by $a$.
We are mainly interested in the region where 
$V_0 \le E_F$ since we wish to study the resonant tunneling regime
and not the CB regime. 

The effective two-channel $S$-matrix  for this system of $n_b$ barriers can 
be written as 
\beq
S = \left( \begin{array}{ll}
re^{i\theta} & f e^{i\phi} \\
f e^{i\phi} & r e^{i\theta'} 
\end{array} \right)
\label{two}
\eeq
where the parameters $r,f,\theta,\theta'$ and $\phi$ are functions
of the Fermi energy $E_F$ and the amplitudes of the time-
varying potentials $V_i(t)$. Their explicit forms can be found, 
in terms of the 
parameters of a single well, (in the adiabatic limit),
by solving the
time-independent Schrodinger equation for the potential $V_i(t)$
given in Eq. \ref{one}, for each value of $t$.
The reflection coefficients  are not the same (phase can differ)
because the time-varying  potentials explicitly violate parity. 
The potential also violates
time-reversal invariance. But since in the adiabatic approximation,
we are only interested in snapshots, at each value of the time, 
the Hamiltonian is time-reversal invariant and hence, the 
transmission amplitudes are the same for the `12' and `21' elements
in the $S$-matrix.

By the Brouwer formula\cite{BROUWER}, the charge pumped
can directly be computed from the parametric 
derivatives of the $S$-matrix. 
For a single channel, it  is given by 
\bea
Q &=& {e\over 2 \pi} \int_0^\tau dt Im {\large (}
{\partial S_{11}\over \partial V_1}S_{11}^* {\dot V_1}
+ {\partial S_{12}\over \partial V_1}S_{12}^* {\dot V_1}
\nonumber \\
&&~~~~~~~~~~\quad+{\partial S_{11}\over \partial V_2} S_{11}^*{\dot V_2}  
+ {\partial S_{12}\over \partial V_2} S_{12}^*{\dot V_2}{\large )}
\eea
where $S_{ij}$ denote the matrix elements of the $S$-matrix
and ${\dot V_1}$ and ${\dot V_2}$ are the time derivatives of
the $V_1,V_2$ given in Eq. \ref{one}.
Thus, the pumped charge is directly related to the amplitudes
and phases that appear in the scattering matrix.
For the form of the $S$-matrix given in Eq. \ref{two}, this is just
\beq
Q = {e\over 2\pi} \int_0^\tau[ {\dot \theta} - f^2({\dot \theta}
-{\dot \phi})] dt
\label{pcharge2}  
\eeq
where the first term is clearly quantised since $e^{i\theta}$ has to 
return to itself at the end of the period. So the only possible 
change in $\theta$ in  a period can be in integral multiples of $2\pi$
-i.e., $\theta(\tau) \longrightarrow \theta(0) + 2\pi n$.
The second term is the `dissipative' term which prevents
the perfect quantisation. It is easy to see the analogy
of Eq. \ref{pcharge2}  with Eq. 19 of Ref.\cite{ALEINER}.

The correlation between resonant  transmission and the pumped charge
has been studied earlier by several groups\cite{WEI,ZHU,AB,EWAA1,EWAA2}. 
A pumping
contour\cite{EWAA1,EWAA2} can be defined in the plane of the parameters,
$V_1$ and $V_2$,
as the closed curve traced out in one cycle of modulation of the
barriers ($t \rightarrow t+\tau$). As we shall see below explicitly,
the value of the first term in Eq. \ref{pcharge2} is set
by the number of transmission maxima enclosed by the pumping
contour, whereas the second term depends on how peaked the
transmission function is, around its maxima. We find that the more
peaked the transmission function, the smaller is the dissipation.

The aim of this paper is to study the effect of electron-electron
interactions on the charge pumped. 
In Ref.\cite{SHARMA}, a bosonised approach was used; here, 
the barriers have to be either in the weak back-scattering or
the strong back-scattering limit. But for time-dependent potentials,
within a single pumping cycle, the barriers can range from the weak
barrier to strong barrier limits, if the 
perturbing potential is large. Contours with such large perturbing
potentials are required for enclosing multiple transmission maxima, which,
in turn, lead to larger pumped charges. 
Hence, it is much more convenient to introduce interactions (provided
they are weak) perturbatively and directly obtain the renormalisation
group (RG) equations for the entries of the $S$-matrix. This method
has been used extensively in Refs.\cite{LAL,DRS} and we directly present the
RG equations for the effective two-channel $S$-matrix given in 
Eq. \ref{two}  as follows - 
\beq
\frac{dS}{dl} ~=~ M ~-~ S M^\dagger S~.
\label{rg}
\eeq
Here $M$ is a diagonal $2\times 2$ matrix given by
\beq
M = \left( \begin{array}{ll}
{1\over 2}\alpha_1 r  & 0 \\
0 & {1\over 2} \alpha_2 r  
\end{array} \right)~
\eeq
and $l$ is the length scale.
The dimensionless constants $\alpha_i$ are
small and positive and are related to the Fourier transform of
the short-range density-density interaction
potential $V_i(x)$ in the two channels as  
\beq
\alpha_i = \frac{({\tilde V}_i (0)- {\tilde V}_i (2k_F) }{2\pi \hbar v_F} ~,
\label{ali}
\eeq
where we assume that the Fermi velocity is 
$v_F = \hbar k_F/m$.
Thus, $\alpha_i$ is a measure of the inter-electron interactions.
It can also be related to the bosonisation parameter $K$ as
\beq
K_i ~=~ \bigl( \frac{1-\alpha_i}{1+\alpha_i} \bigr)^{1/2} ~.
\eeq

When the barriers are symmetric (i.e., at some point in the pumping 
cycle when $V_1 = V_2)$), parity is not violated and we can set 
$\theta =\theta'$ in Eq. \ref{two}. We will always run the RG equations at this
point.  For the computation of the pumped charge, the origin in $t$ is
not relevant. We only need to compute the charge pumped in one cycle.
It is also natural to expect the
two channels to have the same inter-electron interactions; hence $\alpha_i =
\alpha$. In that case, we obtain the
following explicit equations for the RG flow of the transmission
(and reflection) amplitudes and phases. We find
\bea
{df\over dl} &=& -\alpha f (1-f^2) ~,\label{rgtone}\\
{d\phi \over dl} &=&0~,
\label{rgtrans}
\eea 
for the flow of the transmission amplitude and phase
and
\bea
{dr\over dl} &=& {\alpha\over 2} r (1-r^2 -f^2\cos 2(\phi-\theta))~, 
\label{rgone}\\
{d\theta  \over dl} &=& {\alpha\over 2}f^2 \sin 2(\phi-\theta) ~,
\eea
for the flow of the reflection amplitude and phase.
(Note that when $\theta\ne \theta'$,
two similar sets of equations can be obtained, both for $\theta$
and for $\theta'$ using  $S_{12}$ and $S_{11}$ for one and 
$S_{22}$ and $S_{21}$ for the other.) 
However, the condition of unitarity on the $S$-matrix in Eq. \ref{two}
implies that $\phi-\theta = \pi/2 + 2n\pi$. This simplifies the
RG flow equations for the reflection amplitude and phase, so that
they become similar to those for the transmission amplitude and
phase, i.e.,
\bea
{dr\over dl} &=& \alpha r (1-r^2) ~,\label{rgtone2}\\
{d\theta \over dl} &=&0~.
\label{rgtrans2}
\eea 
The entries of $S$ therefore become functions
of the length scale $l$ or equivalently of $L$,
where  $l = {\rm ln} (L/d)$, and  $d$ is the short-distance cutoff which 
we take to be the  
average interparticle spacing. The RG flow can also be considered as 
a flow in the temperature since the  length scale $L$ can be  
converted to a temperature $T$ using   
\beq
L_T ~=~ \frac{\hbar v_F}{k_B T} ~,
\label{lt}
\eeq
as the thermal length. The interpretation of $L_T$ is that it is the
thermal length beyond which the phase of an electron wave-packet 
is uncorrelated. Thus $L_T$ is the thermal phase coherence length.
Hence, the RG flow has to be cutoff by either $L_T$ or the system
size $L_s$, whichever is smaller\cite{DRS}. In all our numerical computations,
we start the RG flow at the microscopic length scale $d$, which is
taken to be the inter-particle separation.
We then  renormalise to larger and larger length scales, stopping 
at either $L_T$ or $L_s$, whichever is smaller.  Note that for a 
length $L=100 \mu m$, the equivalent temperature as computed from Eq.
\ref{lt} is $T=6.5mK$ (for $v_F=3\times 10^7 cm/sec$, which is the
typical value of the Fermi velocity in $GaAs$).
Both of these are experimentally achievable.
Hence, whether $L_s$ or $L_T$ is reached first  depends on
the given system. If $L_T>L_s$, we can continue the RG flow until
we reach $L_T$, where phase coherence is lost due to thermal fluctuations.

In the rest of the paper, we 
compute the transmissions, the phases and  the  quantised charges 
as a function of the various parameters
in the theory. We then show how the elements of the $S$-matrix, and
consequently, the transmissions, the phases and the pumped charge
change as a function of the RG flow. Since the RG flow is also
a flow in temperature, we show
how the pumped charge changes as 
we decrease the temperature
and reaches quantised values at very low temperatures.
However, note that the RG flow is cutoff by the physical size
of the system at a temperature $T_{L_s} ={\hbar v_F}/{k_B L_s}$ and one
cannot go to lower temperatures.  However, one can still continue
the RG flow upto the length scale $L_T$, until the system loses
thermal phase coherence. 
( For illustrative purposes, though, we sometimes take the limit
that  the system size goes to infinity. In that case, the temperature
can be lowered to $T=0$.) We will see that as we change the temperature,
the pumped charge shows power law behaviour as a function of the temperature
until we reach the system size. Beyond that, the pumped charge shows power
law behaviour as a function of the system size.
In the next few subsections,
we explicitly study the double barrier
system $n_b=2$, the triple barrier system $n_b=3$ and the
quadruple barrier system $n_b=4$. We  then extrapolate the results
to an arbitrary number of barriers.  

Strictly speaking, to remain within adiabatic approximation under which the 
Brouwer formula is derived, the energy level spacing in
the dots $\Delta$ has to be larger than the energy scale defined
by the frequency of the time-varying parameter $E_\omega = \hbar
\omega$. It is only under this approximation that the snapshot
picture of studying the static $S$-matrix for different time 
points within the period is valid. A better  approach to go beyond 
the adiabatic approximation\cite{B,WANG}
is to use the Floquet states. Here we remain within the 
adiabatic approximation even though $E_F \ge V_0$, by choosing
the spacing between the barriers $a$ to be sufficiently small,
so that the separation of the resonance level peaks ($\Delta > \hbar\omega$).

\vspace{0.2cm}

\noindent $\bullet$  {\it Single dot case or $n_b=2$} :

\begin{figure}[htb]
\vspace{-1.0cm}
\begin{center}
\epsfig{figure=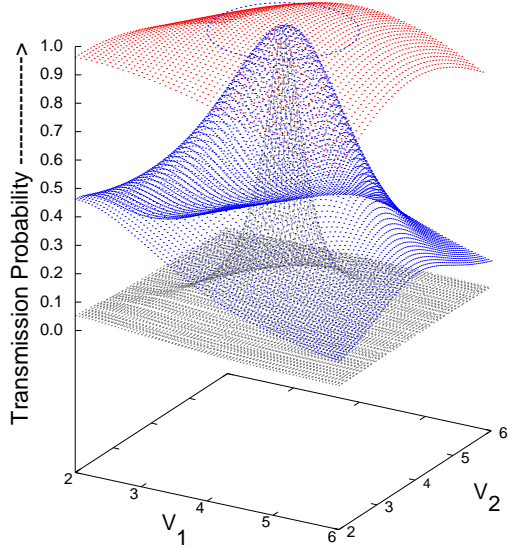,width=6.5cm}
\end{center}
\caption{Transmission $F=f^2$  as a function of the barrier strengths $V_1$
and $V_2$ for three different values of $L=1,100,1000$ 
($T_L = 18 K, 0.18 K, 18mK$)  for the  double
barrier system.  For 
$V_0=4.0, V_p=1.0$, the pumping contour is shown as a dashed line.
The other parameters are given by $E_F = 8.4, \alpha=0.3,
\omega =1$, $a=4$ and $\delta =\pi/2$. The pumped charges are 
$ 0.027, 0.29,0.62$ for the three values of $L$ respectively.}
\end{figure}

Here, we compute the scattering matrix for two $\delta$-function 
barriers  at a distance $a$ apart. 
To facilitate the generalisation to a larger number of barriers,
we first obtain the $M$-matrix linking the incident wave 
to the left of a $\delta$-function barrier  at $x=a$ 
to the transmitted wave to the right of the barrier, in
terms of the strength $V$ of the potential, the distance $a$
and the energy of the incident wave $E_F$ as\cite{AB}
\beq
M = \left( \begin{array}{ll}
1+{i V\over 2\sqrt{E_F}} & ~~{iV\over 2\sqrt{E_F}}e^{-2ia\sqrt{E_F}}  \\
 -{iV\over 2\sqrt{E_F}}e^{2ia\sqrt{E_F}} & ~~1-{i V\over 2\sqrt{E_F}} 
\end{array} \right)~.
\label{m}
\eeq
The $S$-matrix which relates the outgoing waves to the
incoming waves is simply obtained by rewriting the
above matrix elements as
\beq
S = \left( \begin{array}{ll}
M_{21}/M_{11}~ & ~~~~~~1/M_{11} \\
~~1/M_{11}~ & ~-M_{12}/M_{11} 
\end{array} \right)~.
\label{s}
\eeq
Since for a series of potential barriers, the transmitted wave
can be simply obtained by multiplying the $M$-matrices, it is
now easy to obtain the $S$-matrix for any number of barriers
and thus identify the parameters $r,f,\theta,\theta'$ and $\phi$
in terms of the parameters of the potential scattering such
as the distance $a$, the potential strengths $V_i$ and the
Fermi energy $E_F$.
For two barriers, we find the elements of the $S$-matrix as
\beq
S = \left( \begin{array}{ll}
A e^{2ia\sqrt{E_F}}/B & ~~~~1/B \\
~~~1/B & A  e^{-2ia\sqrt{E_F}}/B
\end{array} \right)~,
\label{s2}
\eeq
where $
A = -i (V_1+V_2e^{2ia\sqrt{E_F}})/2\sqrt{E_F} - V_1V_2 
(e^{2ia\sqrt{E_F}} -1)/4E_F$ and  
$B = 1+i(V_1+V_2)/2\sqrt{E_F} +V_1V_2
(e^{2ia\sqrt{E_F}} -1)/4E_F$
and $V_1$ and $V_2$ are the strengths of the two barriers.  

\begin{figure}[htb]
\begin{center}
\epsfig{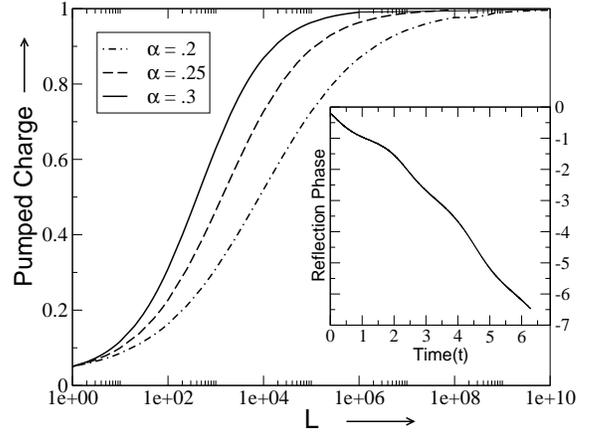}
\end{center}
\caption{$Q$ (in units of $e$) as a function of the RG length scale, 
$L$
for three different values of the interaction strength $\alpha = 0.2,0.25,0.3$
for the double barrier system.
The values of the other relevant parameters are the same as in Fig. 2.
The inset
shows the change in the reflection phase in one pumping cycle to be 2$\pi$.}
\end{figure}

\vspace{0.2cm}
\begin{figure}[htb]
\begin{center}
\epsfig{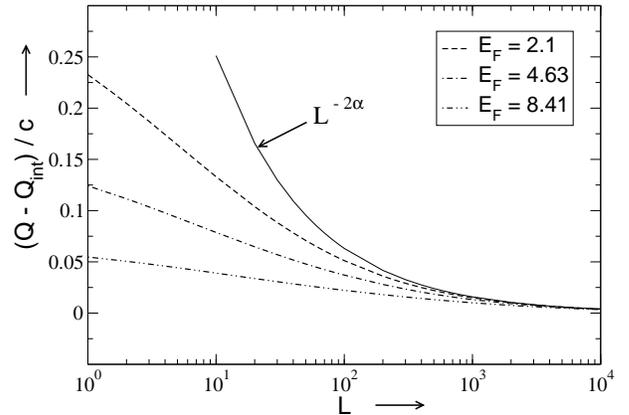}
\end{center}
\caption{Scaling of the dissipative term $Q-Q_{\rm int}$ as a function
of the length scale for the double barrier system. 
As seen in Eq. \ref{qq}, for long length scales,
the term that appears in the denominator can be ignored and the dissipation
terms for different values of $E_F$ fall on the scaling curve $L^{-2\alpha}$
shown in the figure as a solid line. The contour chosen has $V_0=4.5$, 
$V_p=4.0$ and $\delta=\pi/4$.}
\end{figure}

Following Ref.\cite{WEI},
to obtain numerical results, we set the width of the well $a=4$ 
(in units of the inter-particle separation $a_{I} = 100 A^o$) and $\omega =1$.
We find, as expected in the adiabatic approximation,  
that our results are independent of $\omega$
and hence $\omega$ can be made as small as we wish.
We note that our  results are periodic in $a$ for
a fixed $E_F$ (see Eq.\ref{m}) and hence
the width of the well can be increased.  However, to remain within
the adiabatic approximation, $a$ is restricted by $\Delta > \hbar \omega$.
Also, for any value of $a$, one can find the
maximum for the pumped charge by tuning $E_F$\cite{WEI,AB}. 
Our energy units are set by $\hbar=2m=k_B=1$, where $k_B$ is the 
Boltzmann constant. So for $GaAs$, with a typical inter-particle 
spacing $a_{I}
=100 A^o$,  and a typical effective mass $m =0.07m_e$, if we wish to set 
$a_{I}=1$,
the energy unit is $E=5.6 meV$, (using $E=\hbar^2/2ma_I^2$) 
which corresponds to a temperature
of $T=65 ^oK$ (using $E=k_B T$).  
The pumped charge for the  double barrier system was obtained earlier
in Refs.\cite{WEI,AB} and shown to be very small.
Here, we shall see that the reason for this is that 
the transmission is not peaked about the transmission maxima; in fact,
it is rather flat.
Hence, there is a large
dissipative term in the pumped charge. We shall also
see how interactions make the transmission more peaked about the
maxima, reduces the 
dissipation and enhances the possibility of quantisation. 
We compute the transmission as a function of 
the barrier heights $V_1$ and $V_2$  and plot it in Fig. 2,
for a range of $V_i$ between 2.0 and 6.0. 
The pumping contour\cite{LEV} is defined as the closed path formed
in the parameter space by the variation of the parameters in one
cycle. Here, the varying parameters are $V_1$ and $V_2$.
For this graphs, we 
choose the mean barrier height to be $V_0 =4.0$ and the perturbation
to be $V_p=1.0$. We also choose $\delta=\pi/2$ to maximise the
pumped charge. (For the triple barrier and quadruple barrier cases,
different pumping contours are chosen depending on what we need to 
illustrate). These parameters, using Eq. \ref{one}, 
define the pumping  contour, shown in the $V_1-V_2$ plane in Fig. 2 as
a dashed line. The top surface in Fig. 2  denotes
the transmission (at the Fermi energy $E_F=8.4$, which is much
above the barrier heights $4.0\pm 1.0$), 
without any renormalisation of the barriers 
due to interactions. It  is essentially flat and only a small part of
the region of high 
transmission is enclosed by the pumping contour. Thus, although
the pumping contour encloses the transmission maximum, which essentially
implies that the first
term in Eq. \ref{pcharge2} is unity, (see inset in Fig. 3, which shows
that the reflection phase change in one pumping cycle is
$2\pi$), the
second dissipative term is very large and the pumped charge is
vanishingly small ($Q=0.027$).

\begin{figure*}[htb]
\begin{center}
\epsfig{figure=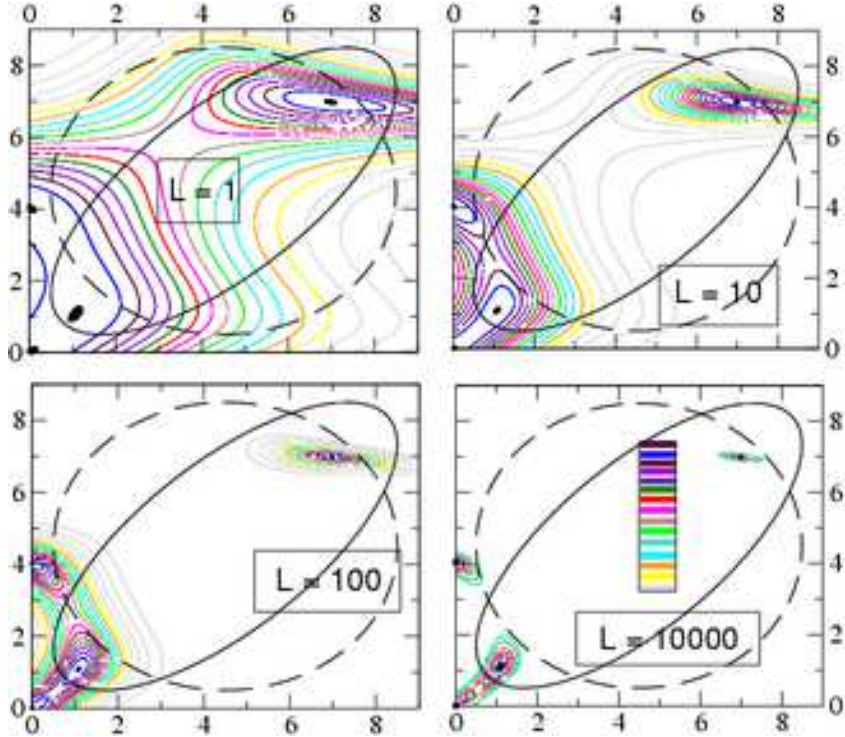}
\end{center}
\caption{Contour plots of the transmission as a function of $V_1$ and $V_2$
for four different lengths $L=1,10,100,10,000$ for
the triple barrier system. The maxima are shown in black
and the separation between the contours is 0.05, whereas the white regions
correspond to transmission less than 0.05. The colour coding for the
transmissions starting from $F=1$ in black to $F=0.05$ in light gray,
is shown in the last panel. The two pumping contours
are for $V_0=4.5$, $V_p= 4.0$ and $\delta =\pi/2$ for the dashed contour
and $\delta=\pi/4$ for the solid contour. The pumped charges
for each of the contours in each of the panels is given in the text.
The other parameters are
as given in Fig.(2).}
\end{figure*}

We now use the RG equations in Eqs.\ref{rgtone}-\ref{rgtrans2}.
Since the phases do not change under the RG flow, we only need
to use Eq. \ref{rgtone}, 
to  compute the change in the
transmission as a function of the length scale. 
We start the RG flow at $L=1=d=4\times 100 A^o$, ($a=4$), which is
the width of the well, and then renormalise to longer length scales,
until we reach the size of the sample, which is at best $100 \mu m$.
At this length scale, the entire sample is phase coherent.
In terms of temperature, the renormalisation is from $T = 18  K$
to $T = 6.5 mK$, which is well within the range of experimental
feasibility. ( However, in many graphs, we show a much larger range
of the renormalisation, assuming that the samples can be made phase
coherent, over much longer length scales, simply for illustrative purposes.)
We find that as the 
RG flow proceeds, the transmission maxima get more and more peaked as
seen in Fig. 2, where we have plotted the transmissions at
different points along the RG flow. The intermediate surface
is when the length scale is $L=100$ or equivalently
$T=.18 K$.  Here, the surface is clearly more
peaked than the original surface and the pumpiong contour encloses
a reasonable fraction of the non-zero transmission. The pumped charge
turn out to be 0.29.
The bottom surface is much
more highly peaked and is the transmission at the length scale $L=1000$
or at $T_L=18 mK$.
Here, clearly, almost all the non-zero transmission is enclosed
by the pumping contour and the pumped charge is 0.62.
 This narrowing of the transmission peaks
is expected from Luttinger liquid studies of the 
double barrier\cite{KF}, and this narrowing is what is responsible for
the fact that the 
dissipative term in Eq. \ref{pcharge2} starts contributing less and less.
As shown in Fig. 3, this leads to charge quantisation at very long length
scales or very low temperatures. Experimentally, it should be possible
to study the pumped charge as a function of the temperature and extract
the interaction parameter $\alpha$ by plotting $Q$ versus $T$. 
In Fig. 3, we have also plotted the change in the pumped charge
as a function of the length scale for three different values of
the interaction parameter $\alpha$. As expected, it reaches close to 
quantisation earliest for the highest value of $\alpha$. We have not taken very
large values of $\alpha$, since this approach to interactions is
perturbative and will not work for strong inter-electron interactions.
For strong inter-electron interactions, one needs
bosonisation\cite{SHARMA}. The pumped charge is perfectly
quantised whenever the backscattering potential leads to insulating
behaviour. The charge quantisation in a double barrier open quantum dot
has also been attributed to Coulomb interactions in Ref.\cite{ALEINER}.
The advantage of our method of introducing interactions
perturbatively is that we can study the crossover from non-interacting
electrons with low values of pumped charge to interacting electrons
with quantised pumped charge at low temperatures. 

We can also explicitly compute how the dissipation term
in  Eq. \ref{pcharge2} scales as we change the length scale (equivalently
temperature). The RG equation for the transmission in Eq. \ref{rgtrans}
can be integrated to yield
\beq
f(l) = {e^{-\alpha l} f_0 \over \sqrt{ r_0^2 + e^{-2\alpha l}f_0^2}}
\label{tl}
\eeq
where $f_0,r_0$ are the values of the transmission and reflection at
$l=1$, or $L=d$, (i.e., in the high temperature limit). Hence at  
length scales $L$ where the second term in the denominator of 
Eq. \ref{tl} can be ignored, we find that $t(L) \propto 
e^{-\alpha l}=(L/d)^{-\alpha}$. 
Using this, we can obtain the scaling behaviour of the pumped charge as
a function of the length scale $L$ or the temperature $T$. In terms
of the Landauer-Buttiker conductance $G_0 = (2e^2/h) f_0^2$, using
Eq. \ref{tl}, we obtain
\bea
Q &=& Q_{\rm int}-{1\over 4\pi e} ({T\over v_F})^{2\alpha} \
\int_0^\tau dt I(t) \nonumber \\
{\rm where}~ I(t)&=&  {G_0(t) {\dot \delta} \over (1-
(G_0/2e^2)\{1-(T/v_f)^{2\alpha}\})^{1/2}}~.
\label{qq}
\eea
$\delta =\theta-\phi$ and 
as earlier, we have used the units $\hbar=k_B=1$.
$Q_{\rm int}$ is the integer contribution of the first term
in Eq. \ref{pcharge2}. Thus, as a function of temperature or the
length scale $L$, the pumped charge scales as 
\beq
Q-Q_{\rm int} = c ({L\over d})^{-2\alpha} = c' T^{2\alpha}~,
\eeq
at low temperatures (or long length scales), where the term that
appears in the denominator of the integrand in Eq. \ref{qq} can
be ignored. 
This can be seen in Fig. 4  where we note that when 
$Q-Q_{\rm int}$ is plotted against $L$, at large values of $L$,
the different curves fall on top of each other.
Note that, as we mentioned before, for length scales
above the system size 
$L_s$, the scaling is no longer in terms of temperature, but
instead is replaced by 
\beq
Q-Q_{\rm int} = c ({L\over d})^{-2\alpha} = c'' L_s^{-2\alpha}~.
\eeq
The same scaling graph is applicable here, since we have given it
in terms of a generic $L$.

\vspace{0.2cm}

\noindent $\bullet$ {\it Double  dot case or $n_b = 3$} :

\begin{figure*}[htb]
\begin{center}
\epsfig{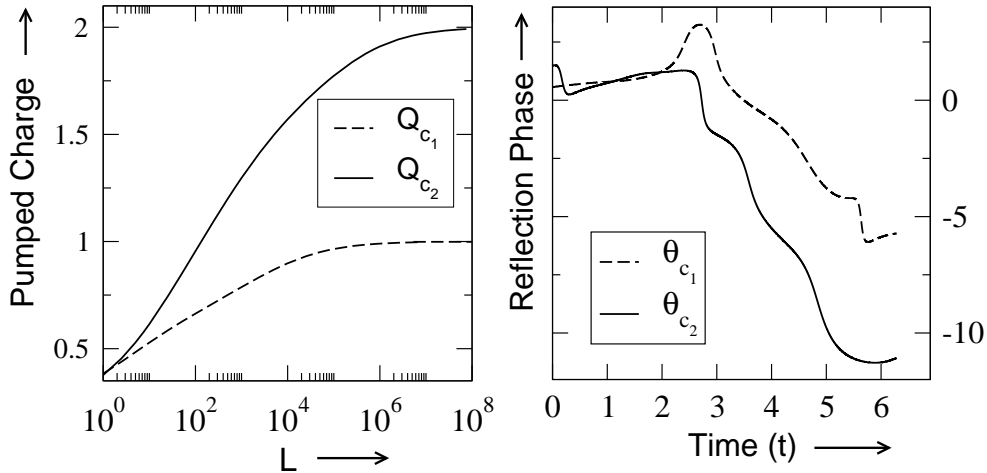}
\end{center}
\caption{$Q$ (in units of $e$) as a function of the RG length scale, 
$L$
for the two different contours shown in Fig. 4 (triple barrier system). 
The appropriate reflection phases
for the two contours are also shown. 
We have
set $V_0=4.5$, $V_p=4.0$, and $\delta =\pi/4$ for $C_1$ and
$\delta =\pi/2$ for $C_2$. We have chosen $E_F=4.6$. The other parameters
are the same as in Fig.(2).}
\end{figure*}

Here, we have computed the $S$-matrix by first obtaining the 
$M$-matrix for $3$ barriers and  obtained the transmission
and reflection coefficients, their phases and the pumped charge.
The contour plots for the 
transmission as a function of the barrier heights $V_1$ and
$V_2$ is given in Fig. 5. The four panels which are for 
four different values of the length
scale, starting with the unrenormalised values of the transmission
for $L=1$, show how the peaking of the transmission occurs as we
go to longer length scales or lower temperatures. As explained
in the figure, the maxima ($T$ very close to unity) are shown in
black. The separation of transmission between the contours is 0.05
and the lowest value of transmission ($T=0.05$) is for the lightest
of the gray contours.     
The new feature that appears here is the presence of more than one 
transmission maxima for a given $E_F$.  
Note that the true maxima appear close to the
$V_1=V_2$ line. This is also true for the double barrier case (see
Fig. 2). The reason is simply because the transmission maximum (=1) can
only be reached for symmetric barriers. Here, the barriers are time-
dependent and are not always symmetric, but the maxima still appear
when the barriers are symmetric. Note also that in Fig. 5, there are
maxima when one of the barrier strengths goes to zero (or one of
the barriers is switched off). This is simply the reflection of the
resonance maximum of the double barrier case and is not relevant for
the genuine triple barrier problem. 
The pumped charge
is now seen to depend on the pumping contour chosen (i.e., the values of the
barrier parameters that are modulated). Two possible pumping contours
are shown in the Fig. 5, with the solid contour $C_1$
enclosing both the transmission maxima and the dashed contour 
$C_2$  only enclosing one. At high temperatures, (first
panel in Fig. 5), the regions of non-zero transmission are spread out
and the  pumped charge can be anything since the dissipation term is large.
It cannot be predicted and there is not much difference in the pumped charge
of the two contours, which are found to be
$Q_{C_1} = Q_{C_2}= 0.38$ for both the curves.
But as we decrease the temperature, we see that the
regions of high transmission are squeezed together and by the time $L=100$,
the two maxima are quite distinct. Both for $L=10$ and for $L=100$,
the solid contour $C_1$ which encloses
more of the areas of non-zero transmission has a higher value of the
pumped charge. For the second panel $L=10$, $Q_{C_1} = 0.73$ and 
$Q_{C_2} = 0.57$ and for the third panel,$L=100$, $Q_{C_1} = 1.12$ and 
$Q_{C_2} = 0.72$. By the time $L=10,000$ (at low temperatures), 
the last panel in Fig. 5  clearly shows that the dashed contour $C_2$ 
encloses all of the transmission around one of the maxima 
and no part of the transmission of the second maxima. Hence, for this
contour, the the charge is quantised
to be almost one ($Q_{C_2} = 0.94$). 
$C_1$, on the other hand, encloses both maxima and most of the non-
zero transmission of the second maxima as well. Here, 
the pumped charge $Q_{C_1} =1.7$,
and further renormalisation is needed for it to reach
the quantisation value of 2.
Thus, for the two barrier case, the pumped charge
can be quantised to be one  or two depending on the pumping contour chosen.
This is also shown in Fig.(6), where we also show the change in 
the reflection 
phase for both contours. Clearly, the phase change is directly related
to the first term in Eq. \ref{pcharge2} , and hence to the charge pumped.

\vspace{0.3cm}
\begin{figure}[htb]
\begin{center}
\epsfig{figure=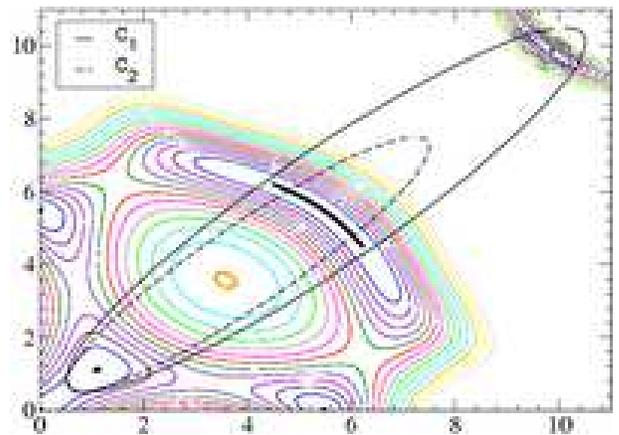,width=8.0cm}
\end{center}
\caption{Contour plots of the transmission as a function of $V_1$ and $V_2$
for four barriers. The conventions are the same as in Fig. 5.
The two pumping contours have $\delta =\pi/8$, with the solid contour $C_1$
having $V_0=5.5$ and $V_p=5.0$ and the dashed contour $C_2$ having 
$V_0=4.0$ and $V_p=3.5$. The reflection phases for these contours are shown in 
Fig.(9).}
\end{figure}

\begin{figure}[htb]
\begin{center}
\epsfig{figure=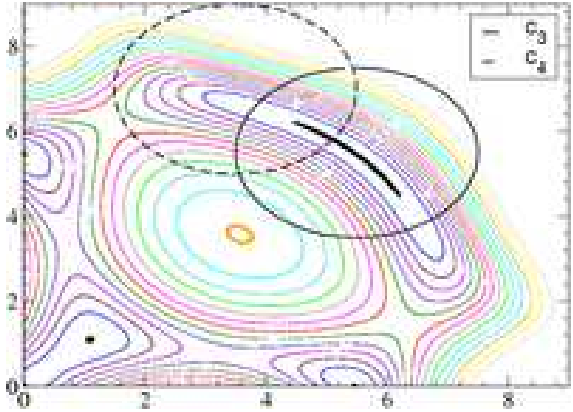,width=7.5cm}
\end{center}
\caption{Contour plots of the transmission as a function of $V_1$ and $V_2$
for four barrier with two pumping contours $C_3$ enclosing the
transmission maximum and $C_4$ not enclosing the transmission maximum.
$C_3$ is defined by $V_0=5.5$ and $V_p=2.0$ and
$C_4$ is defined by $V_{01}=3.5, V_{02}=7.0$ and $V_p=2.0$. 
$\delta=\pi/2$ for both contours.}
\end{figure}

\vspace{0.2cm}

\noindent $\bullet$ {\it Triple dot case or $n_b = 4$} :

Here again, we compute the transmission, the phases and the pumped
charge and the contour plots of the transmission are shown in Figs.
7 and 8. 
The main new feature that appears here is the unusual shape
of the resonance maxima - i.e., in one case, it is almost flat in one direction
and appears as a line. Also, here, we find that there are three
(non-trivial, genuine, quadruple barrier) transmission maxima, which
appear along the $V_1=V_2$ line. (The reason for this has
been mentioned earlier. Also, as has been earlier seen, the other maxima 
that occur for either
$V_1=0$ or $V_2=0$ and are the maxima through triple or double barriers.) 
The charge pumped is a function
of the contour chosen. We can choose a contour that encloses all three
maxima, ($C_1$ in Fig. 7) or one that only encloses two transmission
maxima ($C_2$ in Fig. 7).  In Fig. (9) , the corresponding phase
change in the reflection phase is plotted and  we
see that depending on whether the contour encloses three or two maxima,
the phase change (and consequently, the charge pumped at low temperatures)
is three  or two units. In Fig. (8), we illustrate yet another important
feature. The pumping contour has to actually enclose the centre
of the resonance line; otherwise, the pumped charge is zero at low 
temoeratures. 
This is seen in Fig. 9, where the phase change for the contour 
that encloses the central point ($C_3$) is shown to be $2\pi$ and the phase
change for the contour that does not ($C_4$) is shown to be zero.

\begin{figure*}[htb]
\begin{center}
\epsfig{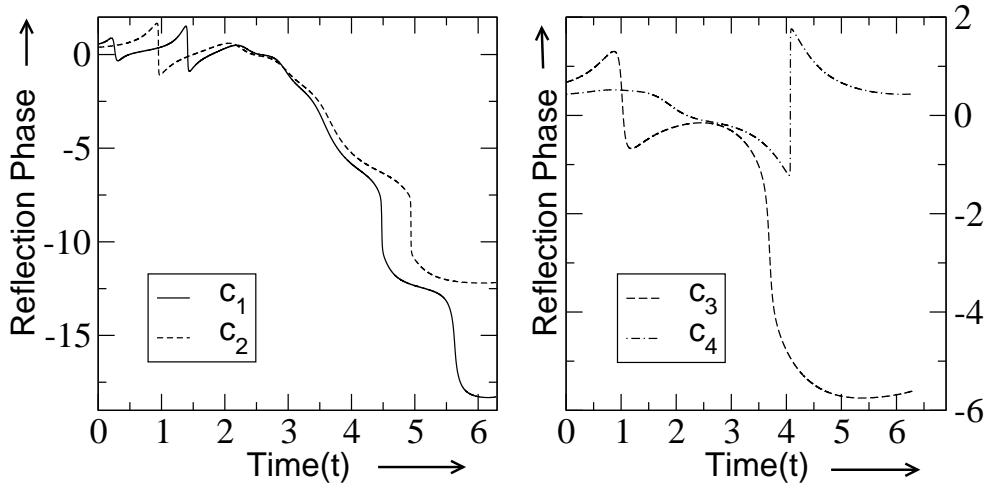}
\end{center}
\caption{The reflection phases as a function of time for the contours
shown in Figs. 7 and 8, (four barrier system). $C_1$ has a phase drop of 
$6\pi$, $C_2$ has a phase drop of $4\pi$, $C_3$ has a phase drop of
$2\pi$ and $C_4$ has no phase drop. These are correlated with quantised
pumped charges of 3, 2, 1 and zero respectively.}
\end{figure*}

These results can easily be extrapolated for the case with arbitrary number
of barriers $n_b$. The results are very similar. For $n_b$ barriers, 
there are $n_b-1$ true maxima close to the $V_1 \sim V_2$ line. If
we can choose a pumping contour that encloses all these maxima, the
drop in the reflection phase will be $(n_b-1) 2\pi$ and the maximum pumped
charge will be $(n_b-1) e$. To obtain quantised values of the charge,
we need to go to low temperatures, so that the dissipative term vanishes.
However, to obtain the value of the pumped charge at low
temperatures, no explicit renormalisation or computation of transmissions
at low temperatures need be performed. We only need to compute
the change in the reflection phase in one pumping cycle for any given contour.
This value is quantised and is a measure of the quantised charge that
can be pumped at low temperatures.

Note that all the qualitative features that we have mentioned above for
the pumped charge for multiple barriers do not depend on the details
of the potentials used, such as the ratio of $V_p$ to $V_0$, or the
value of $E_f$ or the value of $a$. We choose $E_F \ge V_i$ to be in the
resonant tunneling limit. Then as we tune the $E_F$ for fixed values of
the potential, the transmission amplitudes show wide maxima with multiple
($n_b-1$) peaks, periodically as a function of $E_F$. Similar features
are seen for different values of $E_F$. Typically, we have chosen $E_F$
values where the pumped charge is maximised. Similarly, as a function
of the barrier separation $a$, the pumped charge shows periodic behaviour,
and we have chosen an arbitrary value of the separation.

Note also that although these results have been obtained for quantum wells,
they are also applicable to `single level' quantum dots, where there is only
one energy level (close to the Fermi energy) playing a role in the pumping. 
In other words, as long as the pumping frequency
is low, so that $\hbar \omega <\Delta$, where $\Delta$ is the energy separation
between the single level and the other levels in the quantum dot, the
above analysis should hold.

To conclude, in this paper, 
we have examined the effects of electron-electron interactions
on adiabatic quantum pumps and have shown that the limit of `optimal'
quantum pumps ( pumps with no dissipation) are realised at low temperatures,
when the effects of interaction drive the dissipative term to zero.
We have obtained the scaling of the dissipative term as a function
of temperature and shown that it scales as $T^{2\alpha}$ vanishing
at $T=0$. Future extensions include the study of interactions on a spin
pump\cite{DR}

\vskip .2 true cm
\leftline{\bf Acknowledgments :} We would like to thank Argha Banerjee for
collaboration in the early part of this work and Suryadeep Ray for
computational help.

\end{document}